\documentclass[aps,prb,preprint,superscriptaddress,showpacs,floatfix,onecolumn,letterpaper]{revtex4-1}
\usepackage{graphicx}
\usepackage{dcolumn}
\usepackage{bm}
\usepackage{braket} 
\usepackage{times}

\begin{document}


\title{Nanoscale manipulation of the Mott insulating state\\coupled to charge order in 1$\emph{T}$-TaS$_{2}$}
\author{Doohee Cho}
\affiliation{Center for Artificial Low Dimensional Electronic Systems, Institute for Basic Science (IBS), 77 Cheongam-Ro, Pohang 790-784, Republic of Korea}
\affiliation{Department of Physics, Pohang University of Science and Technology (POSTECH), Pohang 790-784, Republic of Korea}
\author{Sangmo Cheon}
\affiliation{Center for Artificial Low Dimensional Electronic Systems, Institute for Basic Science (IBS), 77 Cheongam-Ro, Pohang 790-784, Republic of Korea}
\affiliation{Department of Physics, Pohang University of Science and Technology (POSTECH), Pohang 790-784, Republic of Korea}
\author{Ki-Seok Kim}
\affiliation{Department of Physics, Pohang University of Science and Technology (POSTECH), Pohang 790-784, Republic of Korea}
\author{Sung-Hoon Lee}
\affiliation{Center for Artificial Low Dimensional Electronic Systems, Institute for Basic Science (IBS), 77 Cheongam-Ro, Pohang 790-784, Republic of Korea}
\affiliation{Department of Physics, Pohang University of Science and Technology (POSTECH), Pohang 790-784, Republic of Korea}
\author{Yong-Heum Cho}
\affiliation{Department of Physics, Pohang University of Science and Technology (POSTECH), Pohang 790-784, Republic of Korea}
\affiliation{Laboratory for Pohang Emergent Materials and Max Plank POSTECH Center for Complex Phase Materials, Pohang University of Science and Technology, Pohang 790-784, Korea}
\author{Sang-Wook Cheong}
\affiliation{Laboratory for Pohang Emergent Materials and Max Plank POSTECH Center for Complex Phase Materials, Pohang University of Science and Technology, Pohang 790-784, Korea}
\affiliation{Rutgers Center for Emergent Materials and Department of Physics and Astronomy, Rutgers University, Piscataway, New Jersey 08854, USA}
\author{Han Woong Yeom}
\email{yeom@postech.ac.kr}
\affiliation{Center for Artificial Low Dimensional Electronic Systems, Institute for Basic Science (IBS), 77 Cheongam-Ro, Pohang 790-784, Republic of Korea}
\affiliation{Department of Physics, Pohang University of Science and Technology (POSTECH), Pohang 790-784, Republic of Korea}


\date{\today}




\maketitle

\textbf{Quantum states of strongly correlated electrons are of prime importance to understand exotic properties of condensed matter systems ~\cite{imada1998metal,dagotto1994correlated,lee2006doping} and the controllability over those states promises unique electronic devices such as a Mott memory~\cite{nakano2012collective}. As a recent example~\cite{stojchevska2014ultrafast}, a ultrafast switching device was demonstrated using the transition between the correlated Mott insulating state and a hidden-order metallic state of a layered transition metal dichalcogenides 1$\emph{T}$-TaS$_{2}$. However, the origin of the hidden metallic state was not clear and only the macroscopic switching by laser pulse and carrier injection was reported. Here, we demonstrate the nanoscale manipulation of the Mott insulating state of 1$\emph{T}$-TaS$_{2}$. The electron pulse from a scanning tunneling microscope switches the insulating phase locally into a metallic phase which is textured with irregular domain walls in the charge density wave (CDW) order inherent to this Mott state. The metallic state is a novel correlated phase near the Mott criticality with a coherent feature at the Fermi energy, which is induced by the moderate reduction of electron correlation due to the decoherence in CDW. This work paves the avenue toward novel nanoscale electronic devices based on correlated electrons.}

The electron motion as represented by the bandwidth $W$ is strongly prohibited by the on-site Coulomb repulsion $U$ and a Mott insulating state develops by the localization of electrons near Fermi energy ($E_{F}$) when the $U$/$W$ exceeds a critical value~\cite{imada1998metal}. In a unique Mott insulator of 1$\emph{T}$-TaS$_{2}$, the correlated insulating state is brought by the spontaneous formation of the CDW order, which substantially reduces (increases) the bandwidth $W$  ($U$/$W$) at $E_{F}$~\cite{fazekas1979electrical,smith1985band}. Due to the entanglement, the Mott transition might be controlled by the CDW order. Indeed, the resistance abruptly decreases by one order of magnitude upon increasing the temperature above $T_{c}\sim220$ K~\cite{di1975effects} 
where the CDW order melts to small domains textured by nearly commensurate (NC) domain wall networks~\cite{wu1990direct}.
 The metallic phase with textured CDW can be generated not only by the thermal excitation but also by the chemical doping~\cite{zwick1998spectral, ang2012real, ang2013superconductivity}, the photoexcitation~\cite{hellmann2010ultrafast,stojchevska2014ultrafast}, the pressure~\cite{sipos2008mott}, the carrier injection~\cite{vaskivskyi2014fast,hollander2015electrically} and the reduction of thickness~\cite{yoshida2014controlling,yu2015gate}. 
 
\begin{figure*}
\includegraphics[width=13.0 cm]{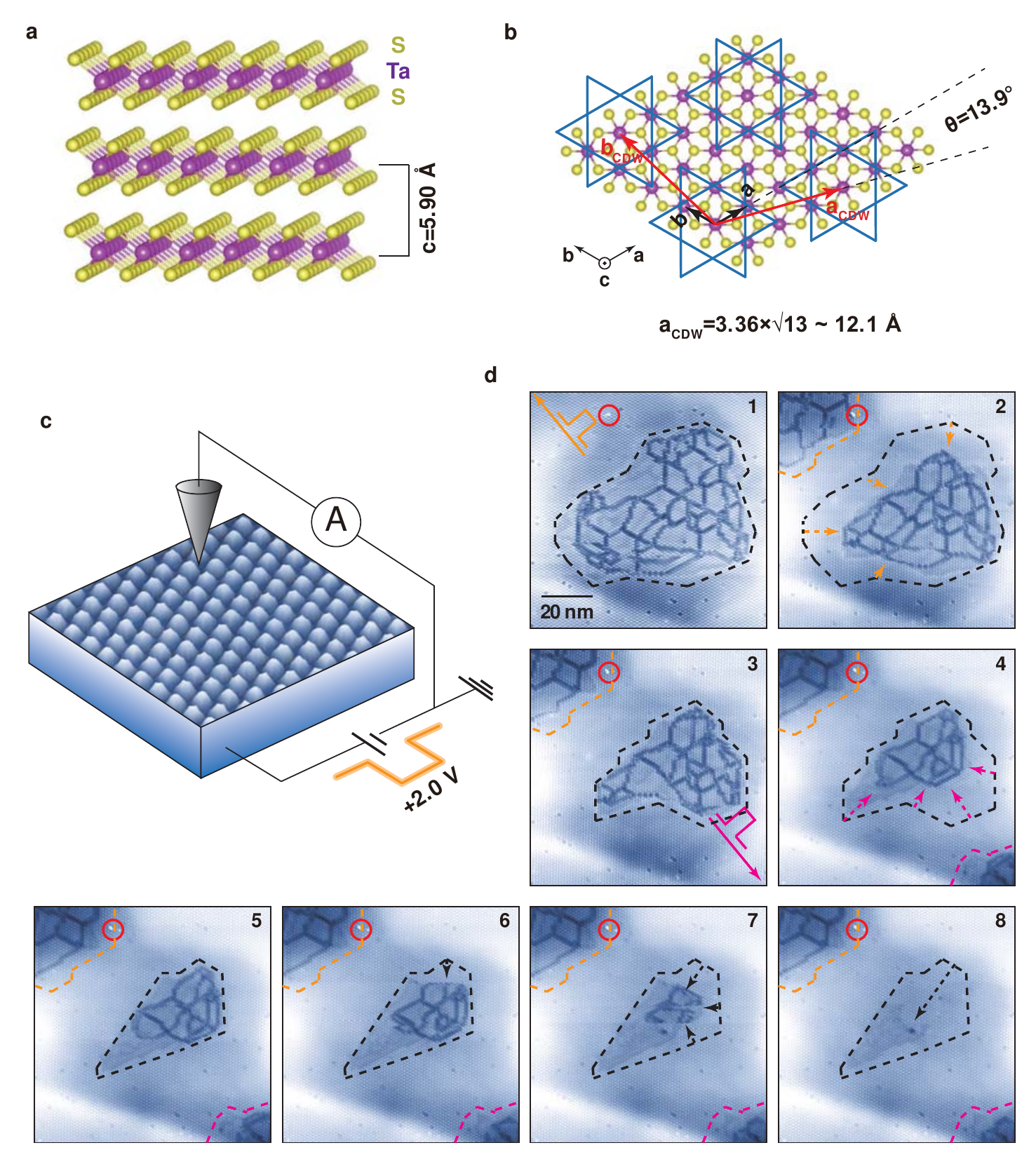}
\caption{\textbf{Nanoscale manipulations via positive voltage pulses to the CDW domain}. \textbf{a,}  Side and \textbf{b,} Top-view of the atomic structure of 1$\emph{T}$-TaS$_{2}$ with the {\it David-star} CDW pattern superimposed. \textbf{c,} STM set-up with an image of the CDW phase (tunneling current $I_{t} =100$ pA, sample bias $V_{s} = -0.80$ V and scan size $L^{2} = 12\times12$ nm$^{2}$) of 1$\emph{T}$-TaS$_{2}$ at $T=4.3$ K. \textbf{d,} A series of STM images ( $I_{t} = 100$ pA, $V_{s} = -1.20$ V and $L^{2} = 92\times92$ nm$^{2}$) showing the time evolution of domains with  the broken phase coherence of the CDW order by the domain wall network. An impurity (red circles) is used as a landmark for all images. Before the STM images of 1, 2 and 4, V-pulses are applied with an amplitude of +2.0 V, +2.10 and +2.35 V with a duration of 100 ms at the central, upper-left and lower-right sites (red and yellow arrows), respectively. The dashed lines and arrows indicate the pulse-induced evolution of domains.}\label{fig1}
\end{figure*}

While these studies demonstrate the macroscopic controllability of the correlated Mott insulating phase by the CDW order, the origin of the excited metallic phase, the textured CDW phase, has been elusive. The metallicity was attributed to the metallic domain wall themselves~\cite{sipos2008mott}, the metallization of the Mott-CDW domains due to the screening by free carriers of domain walls~\cite{zwick1998spectral}, or the change of the interlayer stacking order \cite{ritschel2015orbital}. However, there has been no experimental verification of these scenarios and no direct information on the electronic structure of the domain wall. The nature of the metallic phase is also in the center of the current debate on the mechanism of the superconductivity emerging at low temperatures~\cite{sipos2008mott,joe2014emergence, ang2012real, ang2013superconductivity}.

In the present study, we have succeeded in the nanoscale manipulation of the metal-insulator transition of the Mott insulating phase of 1$\emph{T}$-TaS$_{2}$. A few tens of nanometer metallic patches can reversibly be formed and erased with an atomically abrupt phase boundary by applying voltage pulse from a scanning tunneling microscope tip. The spectroscopy measurements with atomic resolution rule out the existence of substantial free carriers along domain walls 
and unveil the novel correlated nature of the metallic phase.

Figure 1a illustrates the CdI$_{2}$-type crystal structure of 1$\emph{T}$-TaS$_{2}$ with Ta atoms octahedrally coordinated by S atoms. A unit layer consists of one Ta layer sandwiched between two S layers. Within the insulating phase at low temperature, 1$\emph{T}$-TaS$_{2}$ develops a long-range ordered CDW accompanied with the {\it David-star} distortion; 12 Ta atoms shrink toward the center Ta atom and S layers swell up along the $c$-axis. Such a deformation forms a commensurate $\sqrt{13}\times\sqrt{13}$ triangular superlattice ($a_{CDW}\sim12.1$ \AA) (Fig. 1b)~\cite{brouwer1980low}. The lattice deformation brings about the charge localization at the center of {\it David-stars}, which is clearly resolved in the STM image of the CDW phase (Fig. 1c)~\cite{kim1994observation}. 

The manipulation of the Mott-CDW phase was realized by applying a positive voltage pulse ($V_{s} \geq +2.0$ V and $\Delta t = 100$ ms) within a typical scanning tunneling microscope (STM) set-up (Fig. 1c and Methods). A pulse creates a textured CDW domain of a few tens of nanometers with an irregular domain wall network inside. The additional pulse (solid arrows in Fig. 1d) can also reduce the size of a preexisting textured CDW domain (dashed arrows in Fig. 1d). The higher a pulse voltage is, the bigger patch is formed. At this temperature, the textured CDW patch can be very much stable but the STM imaging with a moderate tunneling current gradually reduces its size (from 5 to 8 of Fig. 1d). This indicates the metastability of the textured CDW phase and that we can reproducibly induce a textured CDW patch on a roughly desired position and erase it. 

\begin{figure}
\includegraphics{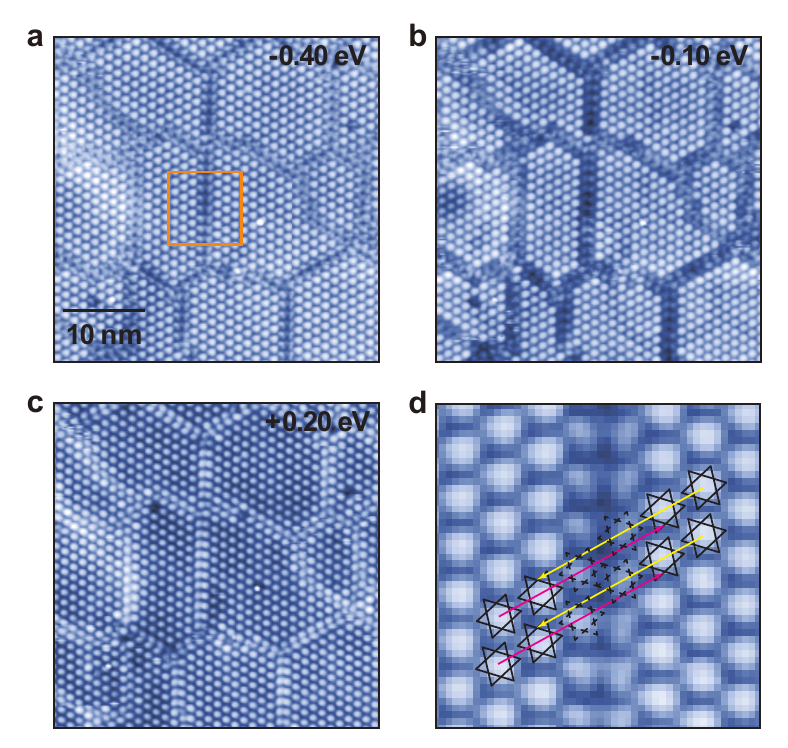}
\caption{\textbf{STM images of the textured CDW phase with domain walls.} \textbf{a}--\textbf{c,} Bias dependent STM images showing the structure and electronic states of domain walls in the textured CDW phase ($I_{t}=100$ pA, $V_{s}=-0.40, -0.10$ and $+0.20$ V and $L^{2} = 40\times40$ nm$^{2}$). \textbf{d,} Zoomed-in of the area in the orange box of Fig. 2a. The phase shift of the CDW ordering across the domain wall is illustrated with the arrows. The dashed {\it David-stars} cannot construct perfect CDW unit cells due to the misfit at the domain wall.}\label{fig2}
\end{figure}

It has been shown that the Mott-CDW phase can be macroscopically turned into the metallic phase with textured CDW by thermal excitation~\cite{di1975effects} or carrier doping~\cite{vaskivskyi2014fast,hollander2015electrically,yoshida2014controlling,yu2015gate}. They commonly introduced the crucial role of extra carriers and, in particular, the temperature-dependent Hall measurement\cite{inada1979hall} and the electric-field-effect study~\cite{hollander2015electrically} indicated the direct role of hole carriers over the critical density in creating topological defects of domain walls. In the present case, only a positive V-pulse is active, where the local hole concentration under the STM tip is enhanced~\cite{brazovskii2014modeling}. This is consistent with the hole-carrier mechanism, while we cannot completely exclude the local heating effect and the transient electron carrier injection by the tunneling current.

The topological defect itself is well resolved in the STM images (Fig. 2). The phase of the CDW changes abruptly across the domain wall as shown in Fig. 2d, which directly affects two adjacent rows of CDW maxima as shown by their reduced contrast. Two rows of reduced CDW maxima and the CDW phase shift is the most common domain wall configuration. While the details were not revealed, this domain wall is consistent with those of the thermally excited NC-CDW phase~\cite{wu1990direct}. However, the NC-CDW phase has ordered Kagome lattices of domain walls~\cite{wu1990direct,spijkerman1997x} while the present domain wall network is disordered. This difference is apparently related to the quenched nature of the present textured CDW patches, being far away from the thermal equilibrium. The present case would be close to the metallic phase induced by the laser excitation from the low temperature phase. This work calls the metallic phase a hidden, thermodynamically unreachable, order state and assumed a triangular lattice of the domain walls without any microscopic information~\cite{stojchevska2014ultrafast}. 

\begin{figure*}
\includegraphics[width=13.0 cm]{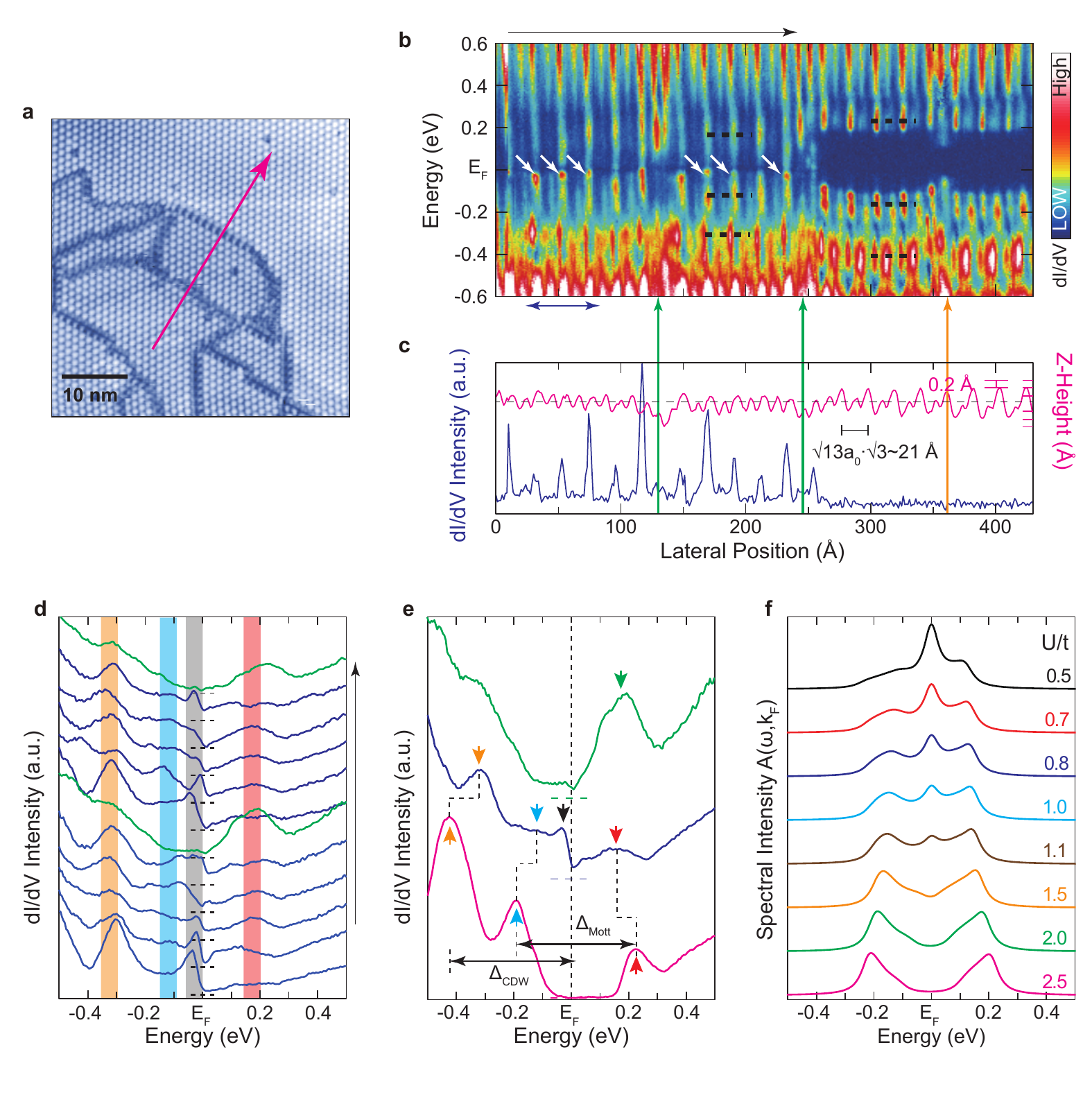}
\caption{\textbf{Direct comparison of electronic structures between commensurate and textured CDW phases} \textbf{a,} STM image of a crossover region between the commensurate and textured CDW phase ($I_{t}=10$ pA, $V_{s}=-1.20$ V and $L^{2}=45\times45$ nm$^{2}$). \textbf{b,} The differential conductance $dI/dV(r,eV)$ data measured along the arrow in Fig. 3a. \textbf{c,} Z-height (magenta) and $dI/dV$ intensity profiles at E$_{F}$ (blue) taken together with the $dI/dV(r,eV)$ data. The vertical orange and green arrows indicate a point defect and domain walls, respectively.  \textbf{d,} $dI/dV(eV)$ curves along the black arrow shown in Fig. 3b. Each curve is averaged within each CDW cluster and shifted for comparison. The spectra in green are for the domain walls in the textured CDW region. Colored vertical bars indicate the prominent peak positions, which are also marked with arrows in Fig. 3e. \textbf{e,} Spatially averaged $dI/dV(V)$ curves of the insulating commensurate CDW phase (magenta), the metallic domain (blue), and the domain wall (green). The horizontal dashed lines mark the zero conductance. Blue (red), yellow and green arrows indicate the lower (upper) Hubbard state, the topmost valence band, and the resonance peak, respectively. \textbf{f,} Theoretically calculated spectral function for different correlation $U/t$ in a triangular lattice.}\label{fig3}
\end{figure*}

Most of the previous works blamed the conducting channel along a domain wall as the origin of the metallic property, directly~\cite{sipos2008mott} or indirectly~\cite{zwick1998spectral}. Nevertheless, there has been very little direct spectroscopic information on domain walls. The present study clearly rules out the dominating metallic nature of the domain wall denying most of the previous metallization scenarios for various textured CDW phases. 
 In the STM image, the contrast of the domain walls is suppressed close to $E_{F}$ (Fig. 2b) and enhanced at $V=+0.20$ eV (Fig. 2c). This indicates that the domain walls have little density of states near $E_{F}$ with its own electronic state at +0.20 eV, which is more directly verified by the spatially resolved $dI/dV(r,V)$ curves (green curves in Figs. 3d and 3e) discussed below.  
 
In stark contrast, the textured CDW patch has a clear metallic characteristic. We took the $dI/dV(r,V)$ curves starting from a textured CDW patch into the normal Mott-CDW background. In the ordered Mott-CDW region, two prominent peaks at $V=-0.19$ and $+0.23$ eV are resolved, which correspond to the lower and upper Hubbard bands, respectively, constituting the Mott gap of $\Delta_{Mott}=0.42$ eV. Beyond the Mott gap, there is additional band splitting away from $E_{F}$, most importantly, around $V= -0.30$ eV. This band gap is known to come from the CDW formation~\cite{smith1985band} and the peak at $-0.42$ eV is ascribed to the top of the valence subbands.~\cite{zwick1998spectral,clerc2006lattice,hellmann20012time} This band splitting makes the necessary condition for the Mott insulating state, \textit{i.e.} a narrow band at $E_{F}$. In contrast, within the textured CDW patch, the tunneling spectra unambiguously indicate finite density of states around $E_{F}$. The zero-bias conductance profile [$dI/dV(r,0)$] in Fig. 3c contrasts sharply the metallic and insulating regions. Note that the zero-bias conductance or the $E_{F}$ density of states is peaked on the CDW maxima within the texture CDW patch. These peaks are due to a pronounced spectral feature very close to $E_{F}$ (white arrows in Fig. 3b, the grey bar in Figs. 3d, and the black arrow in Fig. 3e). In addition, the tunneling spectra in the textured CDW phase exhibit broad features centered at -0.12 and +0.16 eV. They are similar to the Hubbard states in the Mott-CDW phase but shifted slightly toward $E_{F}$ with a substantial reduction of the intensity and a noticeable broadening. On the other hand, the top of the valence band is also shifted substantially toward $E_{F}$ (Fig. 3e).

The spectral characteristics of the textured CDW phase within the Mott gap can be straightforwardly related to the breakdown process of a Mott insulating state. The present system has been effectively described by an one band Hubbard model on a triangular lattice at the half filling (see Methods). Our own theoretical calculation based on spin-liquid physics~\cite{lee2005u1} (see Methods) reveals a metal-insulator transition as a function of $U/t$ ($t$, an intersite hopping integral proportional to $W$), which captures all the major experimental findings; weakening and broadening of the Hubbard states together with the reduction of the Mott gap (Fig. 3f) and the appearance of the coherent resonance peak near $E_{F}$. The coherent peak was assigned as the Abrikosov-Suhl resonance in the previous dynamic mean field theory calculation~\cite{aryanpour2006dynamical}. These theories assure that the present textured CDW phase is a correlated metallic state close to the critical regime of the Mott transition ($U$/$t$ $\sim$1.4 in Fig. 4f). 

What remains to be explained is the origin of the reduced correlation $U$/$W$ to drive the transition. The screening by free carriers of domain walls~\cite{zwick1998spectral} is not likely due to the lack of substantial free carriers in the $dI/dV$ data. Instead, we note that the long range CDW order is lost within the textured CDW patch. The reduced CDW order would naturally decrease the CDW band splitting, which is clearly evidenced by the substantial energy shift of the split valence band top in Fig. 4e. The effect of the reduced CDW band gap on the Mott state can be traced by calculating the evolution of the bandwidth at $E_{F}$. Our own calculations unambiguously show that the bandwidth at $E_{F}$ is linearly increased by the decrease of the CDW order parameter (see supplementary information). The bandwidth increase of the Hubbard bands is evident in the experiment. This is close to the concept of the bandwidth-controlled Mott transition discussed previously~\cite{sipos2008mott,perfetti2003spectroscopic}. In addition, we can suggest that the domain walls act also as the disorder in the Mott system, which generally reduces the correlation energy $U$~\cite{lahoud2014emergence}. Thus, we conclude that the reduced CDW order induces the increased bandwidth $W$ and at the same time reduced $U$, which drives the Mott-CDW state into the critical regime of a correlated metallic state. 

\begin{figure}
\includegraphics{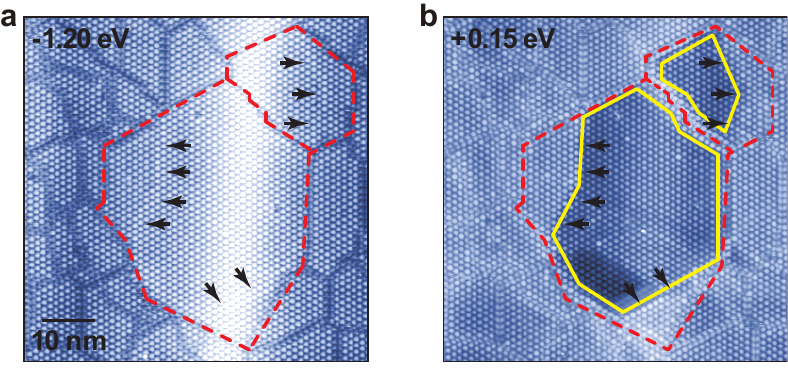}
\caption{ \textbf{Existence of the subdomains and the stacking order} High (\textbf{a,}) and low (\textbf{b,}) bias STM image of the textured CDW domain ( $I_{t}=10$ pA, $V_{s}=-1.20$ and +0.15 V and $L^{2}=90\times90$ nm$^{2}$). The relatively large CDW domains (red lines) within a texture CDW patch and the insulating subdomains (yellow) within them. The black arrows indicate the domain walls in the sublayers.}\label{fig4}
\end{figure}

While the STM images we showed so far only deal with the lateral CDW ordering, we also notice that the vertical CDW stacking order has a crucial impact on this transition. Figure 4 shows that a relatively large domain within the metallic patch has extra but weak domain walls inside (black arrows in Fig. 4a), which is assigned as the domain wall existing in the sublayer(s). The low bias images and the spectroscopy data indicate clearly that this domain harbors insulating subdomains as defined by the sublayer domain walls. One can straightforwardly deduce that these subdomains correspond to different interlayer stacking of the CDW; the interlayer CDW stacking order is lost by the formation of the irregular domain wall networks in each layer except for those metallic subdomains. This suggests that not only the intralayer but also interlayer CDW order has to be taken into account to explain the reduced electron correlation for the Mott transition. The importance of the interlayer CDW stacking order was recently discussed while the direct experimental information has been lacking~\cite{ritschel2015orbital}.

The present case of 1${T}$-TaS$_{2}$ is one of the very rare example of the nanoscale control over strongly correlated electronic states. We can find only one case for the nanoscale manipulation of a Mott insulator in GaTa$_3$Se$_8$~\cite{dubost2013resistive}. In this case, the strong electric-field-induced deformation of the lattice occurs and the transition is close to the avalanche dielectric breakdown~\cite{guiot2013avalanche}. In the present system, the transition is largely electronic and the resulting metallic state is a novel correlated state near the Mott criticality. The uniqueness of the present system lies on the intercoupled nature of the Mott state with the CDW order, which provides the extra tunability of the Mott state by a distinct order. The ultrafast switching capability of the present system~\cite{stojchevska2014ultrafast,vaskivskyi2014fast,hollander2015electrically} in combination with the nanoscale controllability is expected to provide a unprecedented novel device platform based on correlated electronic systems.\newpage

\noindent\textbf{Methods}\newline
\noindent\textbf{Preparation of  single crystal 1\emph{T}-TaS$_{2}$.} The single crystals 1$T$-TaS$_{2}$ were grown by iodine vapor transport method in the evacuated quartz tube.  Prior to growth of the sample,  the powder 1\emph{T}-TaS$_{2}$ was sintered for 48 hours at 750 $^{\circ}$C. We repeated this process two times to get poly-crystals. In order to get high quality sample, the seeds were slowly transported by iodine at 900$\sim$970 $^{\circ}$C for 2 weeks. The tube was rapidly cooled down to room temperature in the air due to the metastability of the 1\emph{T} phase.\newline

\noindent\textbf{STM and STS measurements.} The STM and STS measurements have been performed with a commercial STM (SPECS) in ultra high vacuum at $T=4.3$ K. The STM tips were prepared by mechanically sharpened Pt-Ir wires. All of STM images are acquired with the constant current mode with bias voltage $V$ applied to the sample with a fixed duration of $t=100$ ms and a varying amplitude. When the positive V-pulses are applied to the sample, the feedback loop is opened with scanning condition ($V=-0.8\sim-1.2$ eV and $I_{t}=10\sim100$ pA). We can get the small texture CDW domain below $V_{p}\sim+2.0$ eV and a larger textured CDW phase with increasing amplitude. The dI/dV(eV) curves are recored using a lock-in technique with voltage modulation $V_{m}=10$ mV and frequency $f=1$ kHz.\newline

\noindent\textbf{Theoretical calculations.}
The CDW-Mott state in 1$T$-TaS$_{2}$ can be described by an effective one-band Hubbard model on the triangular lattice at half filling 
\begin{equation} \label{Ham1}
H = - t \sum_{\Braket{ij} \sigma }  ( c_{i \sigma}^{\dagger} c_{j \sigma} +  H.c.)
- \mu \sum_{i \sigma } c_{i \sigma}^{\dagger} c_{i \sigma}
+ U \sum_{i} n_{i \uparrow} n_{i \downarrow},
\end{equation}
where $c_{i, \sigma}$ is an electron annihilation operator with spin $\sigma$ at site $i$, identified with each center of {\it David-star}. Hinted from the observation that the coherent peak emerges in the textured CDW state, we take the U(1) slave-rotor representation for possible spin-liquid physics at least in the intermediate temperature regime~\cite{florens2004slave}. It is straightforward to perform the saddle-point analysis based on an ansatz for spin liquid physics~\cite{lee2005u1}, which leads us to confirm the existence of a metal-insulator transition from a spin-liquid-type Mott insulating state to a correlated metallic phase at critical value of $U$/$t$ (see Supplementary information for details). \newline

\noindent\textbf{Acknowledgments}\newline
This work was supported by the Institute for Basic Science (Grant No. IBS-R014-D1). YHC and SWC are partially supported by the Max Planck POSTECH/KOREA Research Initiative Program (Grant No. 2011-0031558) through NRF of Korea funded by MEST. SWC is also supported by the Gordon and Betty Moore FoundationÕs EPiQS Initiative through Grant GBMF4413 to the Rutgers Center for Emergent Materials.\newline
\newpage

\begin{center}
\textbf{\large Supplementary Information for ``Nanoscale manipulation of the Mott insulating state coupled to charge order in 1\emph{T}-TaS$_{2}$''}
\end{center}
\newpage

\section{Creation and Annihilation of the Textured CDW Domain}
 In this section, we explain the details of the nanoscale manipulation in 1\emph{T}-TaS$_{2}$. 
 At first, we acquired STM image of the commensurate CDW domain with several intrinsic defects which can be a landmark (Fig. S1a). After applying a positive voltage pulse ($V_{pulse}\ge+2.0$ V and $t=100$ ms) to the well ordered Mott-CDW state, the creation of the textured CDW patch is confirmed by acquiring STM images (Fig. S1b). The contrast of the textured domain is suppressed and the surface and subsurface domain walls are simultaneously generated. The minimum voltage required is about + 2.0 V, where the smallest textured CDW patch as shown in Fig. S1a is created. The size of the patch increases as the pulse amplitude grows. No patch is formed with negative voltage pulses. The annihilation of the metastable phase with the domain walls, then, can be induced by applying another V-pulse near the textured domain or scanning over it repeatedly (Fig. S1c). The hidden states generated by macroscopic perturbations in 1\emph{T}-TaS$_{2}$ have been known to be annihilated by the thermal annealing process ($T\sim70$ K) due to its prominent stability~\cite{stojchevska2014}.
   
\begin{figure*}
\renewcommand{\thefigure}{S1}
\includegraphics{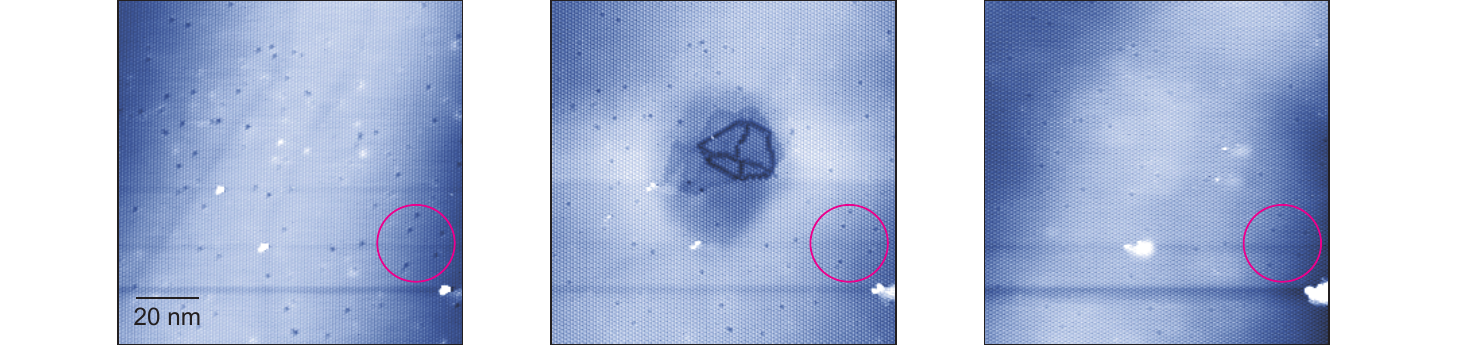}
\caption{\textbf{Creation and annihilation of a small textured CDW Domain.} \textbf{a,} STM image ($I_{t}=10$ pA, $V_{S}=-0.80$ V, $L^{2}=110\times110$ nm$^{2}$) intrinsic several defects (a land mark indicated by a magenta circle) \textbf{b,} Creation of the textured CDW domain by applying V-pulse (V$_{pulse}=+2.0$ V and $t=100$ ms).  \textbf{c,} Annihilation of the metastable state by scanning. Although the tip condition was changed by applying V-pulse process, the quality of the STM images is enough to demonstrate the nanoscale manipulation.}\label{figs1}
\end{figure*}

\section{Dependence of the bandwidth on the CDW order}

In order to investigate the relationship between the CDW order and the bandwidth of the Mott band of the CDW-Mott insulator 1{\it T}-TaS$_2$, we performed density-functional theory calculations that employ the generalized gradient approximation (GGA) \cite{Perdew1996} and the projector-augmented wave method \cite{Blochl1994}, as implemented in VASP \cite{Kresse1996,Kresse1999}. 
Valence electron wavefunctions were expanded in a plane wave basis set with a cutoff energy of 259 eV. The $k$-point integration was performed using a uniform with a $4\times4$ mesh for the Brilloin zone of the $(\sqrt{13}\times\sqrt{13})$ cell with the {\it David-star} distortion (Fig.~S2~a). 

The star distortion makes the broad metallic band split into several subband manifolds with a narrow half-filled band at the Fermi level (Fig.~S2~b).
There are two important energy scales to characterize the electronic structure of the CDW state, the bandwidth ($W$) of the narrow band and the CDW gap ($\Delta_{\rm CDW}$). The latter defined as the energy gap between the Fermi level and the edge of the lower subband reflects the extent of CDW order.
The dependence of the bandwidth on the CDW order was examined by calculating $W$ and $\Delta_{\rm CDW}$ as a function of the CDW order parameter, the size of the star distortion, using the linearly interpolated structures of the undistorted $(1\times1)$ and the fully relaxed $\sqrt{13}\times\sqrt{13}$ structure (Fig.~S2~c).
At the full relaxation, $W$ and $\Delta_{\rm CDW}$ were calculated to be 30 meV and 202 meV, respectively.
When the star distortion diminishes, the bandwidth increases, while the CDW gap decreases, establishing a clear inverse proportion between them, in consistent with the earlier expectations \cite{sipos2008,perfetti2003}.
The present result demonstrates that the CDW order can give us a extra controllability for the bandwidth-controlled metal-insulator transition in CDW-Mott insulator 1$T$-TaS$_{2}$.

\begin{figure*}
\renewcommand{\thefigure}{S2}
\includegraphics{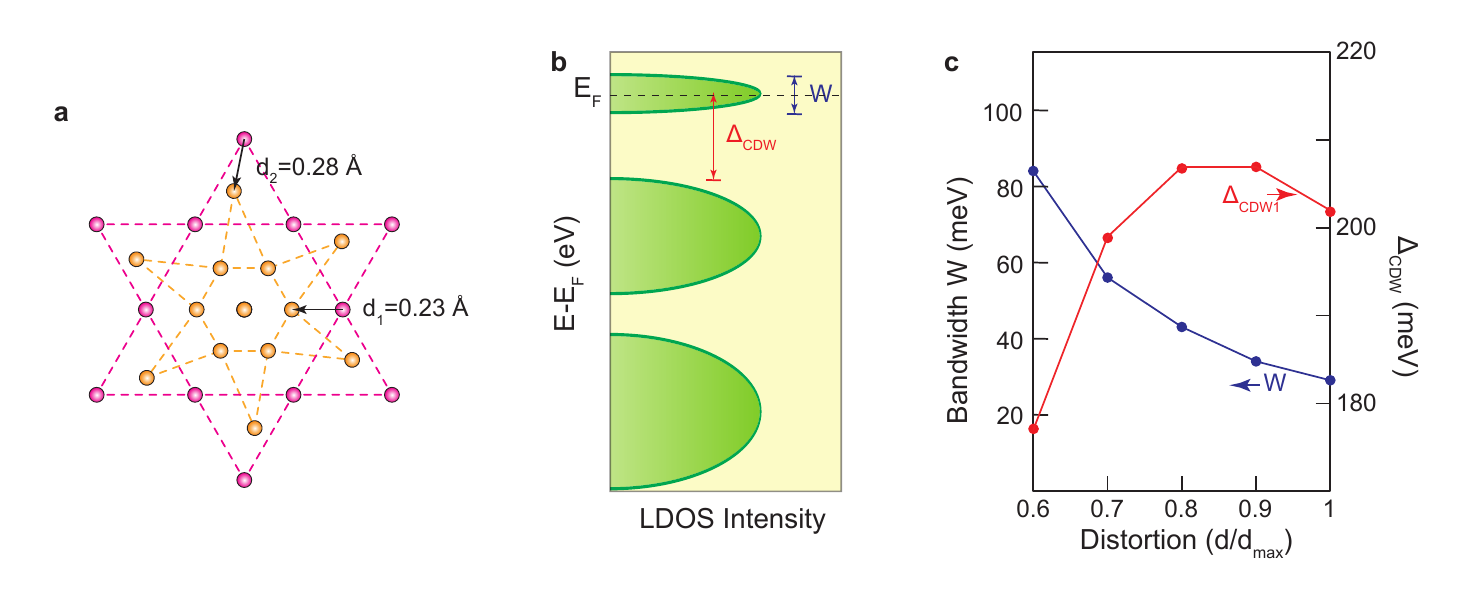}
\caption{\textbf{Dependence of the narrow-band bandwidth on the CDW order} \textbf{a,} The atomic structure of the $1\times1$ triangular lattice (magenta dashed lines) and the {\it David-star} distortion (orange dashed lines). The atomic displacements are exagerated.  \textbf{b,} The schematic view of the CDW-induced band splitting. \textbf{c,} The evolution of the bandwidth and the CDW gap as a function of the degree of the star distortion. The atomic structures of intermediate distortions are taken by linearly interpolating the undistorted $(1\times1)$ and the fully relaxed $\sqrt{13}\times\sqrt{13}$ structure}\label{fig2}
\end{figure*}

\section{Caculation details for a metal-insulator transition in a spin-liquid-type Mott insulator}
The correlations between unpaired electrons on the center of {\it David-star} seem to be responsible for the electrically controlled metal-insulator transition. It can be described by an effective one-band Hubbard model on the triangular lattice at half filling
\begin{eqnarray} \label{Ham2}
H = - t \sum_{\Braket{ij}, \sigma }  ~( c_{i \sigma}^{\dagger} c_{j \sigma} +  H.c.)
- \mu \sum_{i, \sigma } c_{i \sigma}^{\dagger} c_{i \sigma}
+ U \sum_{i} n_{i \uparrow} n_{i \downarrow} .
\end{eqnarray}
$c^{\dagger}_{i, \sigma}$ ($c_{i, \sigma}$) is an electron creation (annihilation) operator with spin $\sigma$ at site $i$. The site $i$ corresponds to the center of {\it David star}. $n_{i, \sigma}$ is the number operator for spin $\sigma$ at the site $i$. $\mu$ is an electron chemical potential, which fits the number of such unpaired electrons at half filling. $U$ is the on-site Coulomb energy and $t$ is the hopping integral between the nearest neighbors.

Resorting to the U(1) slave-rotor representation $c_{i \sigma} = e^{-i \theta_i } f_{i \sigma}$ for possible spin-liquid physics at least in the intermediate temperature regime~\cite{lee2005},
where bosonic field $\theta_i$ describes dynamics of collective charge fluctuations (sound modes) and $ f_{i \sigma}$ expresses a fermionic field for spin degrees of freedom~\cite{florens2004}, we reconstruct an effective theory from the Hubbard model  in terms of such bosonic and fermionic fields. As a result, it is given by
\begin{eqnarray}
\hspace{-0.5cm}
S_F & = &\int_{0}^{\beta} d\tau
\left [
\sum_{i, \sigma}  f_{i \sigma}^{\dagger} (\partial_{\tau} - \mu ) f_{i \sigma}
 - t \chi_f \sum_{\Braket{i j}, \sigma}  ( f_{i \sigma}^{\dagger} f_{j \sigma} + H.c.)
\right ], \\
\hspace{-0.5cm}
S_B & = & \int_{0}^{\beta} d\tau
\left [
\frac{1}{2U} \sum_{i} (\partial_{\tau} b_i^{\dagger}) (\partial_{\tau} b_i)
- t \chi_\theta \sum_{\Braket{i j}} ( b_{i \sigma}^{\dagger} b_{j \sigma} + H.c.)
+ \lambda \sum_{i} (|b_i|^2-1) + 2 L^2 z t \chi_f \chi_{\theta}
\right],
\end{eqnarray}
where the conventional saddle-point approximation has been performed for a spin-liquid-type Mott insulating phase. Here, $\chi_f$ and $\chi_\theta$ describes band renormalization for electrons and the width of incoherent bands, respectively. $\lambda$ is a Lagrange multiplier field to control the spin-liquid to Fermi-liquid phase transition, regarded as the chemical potential of bosons. These equations are based on a nonlinear $\sigma$-model description, where the rotor variable $e^{- i \theta_{i}}$ is replaced with $b_{i}$ and unimodular constraint $|b_{i}|^{2} = 1$ is taken into account~\cite{florens2004}. $z=6$ is the nearest coordinate number of the triangular lattice. $L^{2}$ is the size of system.

\begin{figure*}
\renewcommand{\thefigure}{S3}
\includegraphics{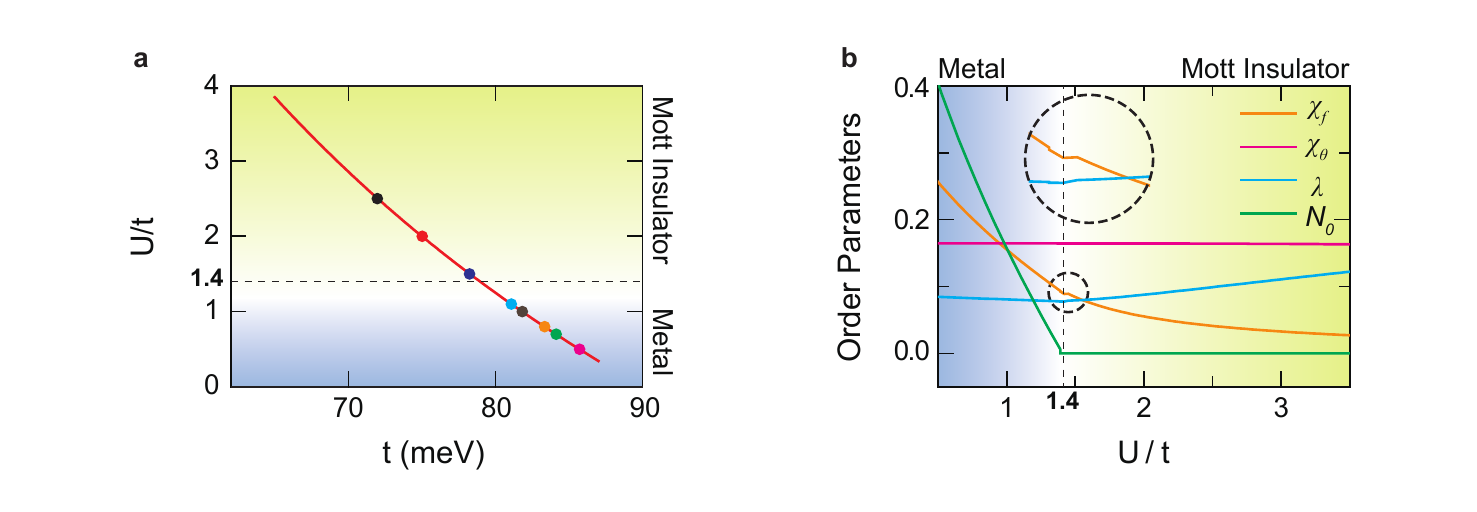} 
\caption{
\textbf{$U$/$t$ controlled metal-insulator phase transition in spin-liquid description.} 
\textbf{a,} The phase diagram and the way to control $U$ and $t$ for spectral functions in Fig. 3f. Each spectral function was acquired at the colored dots.
\textbf{b,} Using the parameters in Fig. S3a, 
the amplitude of order parameters $\chi_f, \chi_\theta, \lambda$, and $N_0$ are calculated self-consistently.
These parameters show the Mott transition at $U/t \approx 1.4$.
The circle-shaped insect shows a kink-like behavior,
which is consistent with the usual phase transition nature in the spin-liquid description~\cite{lee2005u1}.
The lattice size is $30 \times 30$ and the temperature is set to be $T=4.3$ K from the experimental condition.}\label{figs3}
\end{figure*}

Performing the Fourier transformation and the Gaussian integration for both bosons and fermions, we obtain the mean-field free energy 
\begin{eqnarray}
F_{MF} & = & F_F + F_B + L^2 (2 z t \chi_f \chi_{\theta} - \lambda), \\
F_F & = & - \frac{N_\sigma}{\beta} \sum_{\mathbf{k}}
\ln \left [  1+ e^{ - \beta E_F(\mathbf{k})   } \right ],
\\
F_B & = & \frac{1}{\beta} \sum_{\mathbf{k}}
\left (
\ln \left [  1 - e^{- \beta E_B(\mathbf{k}) } \right ]
+ \ln \left [e^{ \beta E_B(\mathbf{k}) } -1 \right ] \right ) ,
\end{eqnarray}
where $ E_F(\mathbf{k})  =    t \chi_f \epsilon_{\mathbf{k}} - \mu$ is the dispersion of fermions and $ E_B(\mathbf{k})  = \sqrt{2 U ( t \chi_{\theta} \epsilon_{\mathbf{k}} + \lambda) }$ is that of bosons. $N_{\sigma} = 2$ represents the spin degeneracy. $\epsilon_{\mathbf{k}}$ is the energy dispersion relation for electrons on the triangular lattice 
at $U=0$ and $t=1$.

Minimizing the effective free energy $F_{MF} = F_{MF}(\chi_f, \chi_\theta, \lambda)$ with respect to $\chi_f$, $\chi_\theta$, and $\lambda$,
we find a metal-insulator transition from a spin-liquid-type Mott insulator to a correlated metal 
at $U/t \approx 1.4$ as shown in Fig. S3b. It is important to note that the $U$ and $t$ can be controlled by the strength of the commensurate CDW ordering as discussed in the section II.

Considering the U(1) slave-rotor decomposition representation, it is straightforward to find that the electron spectral function is given by the convolution integral between fermion and boson propagators, 
\begin{eqnarray}
G(\mathbf{k}, i \omega) 
&=& \frac{1}{\beta} \sum_{i \Omega} \int \frac{d^2 \mathbf{q}}{ (2 \pi) ^2 }
G_F (\mathbf{k+q}, i \omega + i\Omega ) G_B (\mathbf{q}, i\Omega ),
\end{eqnarray}
where 
\begin{equation}
G_F( \mathbf{k}, i\omega) =
\left[ i \omega + \mu - t \chi_f  \epsilon_{\mathbf{k}} \right ]^{-1}, ~~~
G_B( \mathbf{k}, i\Omega) =
\left[  \frac {\Omega^2}{2 U} + t \chi_{\theta} \epsilon_{\mathbf{k}} + \lambda  \right ]^{-1}
\end{equation}
are fermion and boson propagators, respectively. Then, the electron spectral function consists of coherent and incoherent parts, given by 
\begin{eqnarray}
A_{\text{incoherent}} (\omega, \mathbf{k = k}_F)
&=&  \int \frac{d^2 \mathbf{q}}{ (2 \pi) ^2 }
\frac{ U }{ E_B }
\left[
\left \{ n_F (E_F)  + n_B (E_B) \right \} \delta (\omega - E_F  + E_B )
\right .
\\
&& ~~~~~~~~~~~~~~~~~~~~~~~~~~~
-
\left.
\left \{ n_F (E_F)  + n_B (-E_B)  \right \} \delta (\omega - E_F  - E_B )\right], \nonumber \\
A_{\text{coherent}} (\omega, \mathbf{k = k}_F)
&=& N_0 \delta (\omega)
\end{eqnarray}
at the Fermi energy. Here, $n_F(x)$ and $n_B(x)$ are Fermi-Dirac and Bose-Einstein distribution functions, respectively. $ E_F  =    E_F(\mathbf{k}+\mathbf{q})$ and $ E_B  = E_B(\mathbf{q}) $ are $\mathbf{q}$-dependent dispersions at $\mathbf{k=k}_F$. $N_0$ is the condensation amplitude of bosons, which determines the height of the coherent peak in the correlated metallic phase ($N_0\neq0$).

As shown in Fig. 3f, an important result of the spin-liquid approach is that the coherent peak in the electron spectral function increases gradually with the decreasing $U/t$ within the correlated metallic regime. And the incoherent part constitutes the double peaks identified with Hubbard bands. They turn out not only to exist inside the Mott insulating state but also to persist rather deep inside the correlated metallic state.\newline

\end{document}